# Millimeter Observations of the Transition Disk around HD 135344B (SAO 206462)

A-Ran Lyo[1], Nagayoshi Ohashi[2], Chunhua Qi[3], David J. Wilner[3], and Yu-Nung Su[2]

[1] Korea Astronomy and Space Science Institute, 61-1, Hwaam-dong, Yuseong-gu, Daejeon 305-348, Korea,

[2] Institute of Astronomy and Astrophysics, Academia Sinica, P.O. Box 23-141, Taipei 106, Taiwan,

[3] Harvard-Smithsonian Center for Astrophysics, 60 Garden Street, Cambridge, MA 021138, USA

arl@kasi.re.kr

# ABSTRACT

We present ~1″ resolution 1.3 millimeter dust continuum and spectral line ($^{12}$CO and $^{13}$CO J=2–1) observations of the transitional disk system HD 135344B obtained with the Submillimeter Array. The disk shows a Keplerian rotation pattern with an inclination of ~11°, based on the spatially and spectrally resolved $^{12}$CO and $^{13}$CO emission. The data show clear evidence for both dust and gas surface density reductions in the inner region of the disk (radius ≲ 50 AU) from the continuum and $^{13}$CO J=2–1 data, respectively. The presence of this inner cavity in both the dust and gas is more consistent with clearing by giant planet formation than by photoevaporation or by grain growth. There is also an indication of global CO gas depletion in the disk, as the mass estimated from $^{13}$CO emission (~$3.8 \times 10^{-4}$ M$_\odot$) is about two orders of magnitude lower than that derived from the 1.3 mm continuum (~ $2.8 \times 10^{-2}$ M$_\odot$).

Subject headings: planetary systems: protoplanetary disks — stars: pre-main sequence — stars: individual (HD 135344B)

## 1. Introduction

Protoplanetary disks surrounding young stars are the favorite sites of planet formation. However, the detailed physical mechanisms of disk evolution and planet formation are still poorly constrained due to the limited spatial resolution of current observations. In particular, the gas and dust clearing processes in protoplanetary disks are among the outstanding issues, because the lifetime of disks is a key parameter in planet formation scenarios. In addition, the disk clearing could be caused by dynamical interaction with young giant planets (e.g. Papaloizou et al. 2007). In order to study the gas and dust clearing process, "transition" disk systems are especially important because they show the signature for gap/hole in the disk from the spectral energy distribution (SED) (Strom et al. 1989). They are considered to be in the intermediate evolutionary stage between gas-rich protoplanetary disk (classical T Tauri star; CTTs) and gas-poor debris disks (weak-lined T Tauri star; WTTs). For example, Andrews et al. (2011) have studied 12 transition disks with observations of resolved submillimeter dust emission to investigate the disk clearing process. Understanding how gas as well as dust depletes in the disk clearing process requires detailed studies of both spectral line and continuum emission.

HD 135344B (SAO 206462) is a ~8-17 Myr old Herbig F star located at a distance of ~140 pc (van Boekel et al. 2005, Grady et al. 2009). It is the companion star of HD 135344A (SAO 206463) with separation of 21″ (~3000 AU at d=140 pc) (Mason et al. 2001). However, it seems like not to possess any more close companions based on the null results of several companion-searching studies (Pontoppidan et al. 2008, Grady et al. 2009, Müller et al. 2011).

This nearby system has been suggested to harbor a rare transitional or gapped disk, based on a dip in infrared excess above the photosphere at 10 μm, while the excess at the longer wavelength (≳ 10 µm) comparable with that expected from the optically thick disk in its SED (Najita et al. 2007, Grady et al. 2009). The relatively high mass accretion rate of $10^{-8.27}$ $M_\odot$ yr$^{-1}$ (Garcia Lopez et al. 2006) supports that it is at the earlier evolutionary stage than the gas-poor debris disk. The analysis of the SED covering from optical to sub-mm wavelength suggests a large inner gap in the disk of radius R~45 AU (Brown et al. 2007). Furthermore, Müller et al. (2011) suggested that the narrow inner ring from 0.05 AU to 1.8 AU is highly inclined compared to the outer disk based on the different inclination results from 11° to 61° by several near-and mid-infrared continuum and CO molecular gas emission studies (Dent et al. 2005, Doucet et al. 2006, Fedele et al. 2008, Pontoppidan et al. 2008, Grady et al. 2009).

Recently, sub-arcsecond submillimeter dust continuum observation with the Submillimeter Array (SMA) at 0.87 millimeters directly imaged the proposed inner, dust-cleared region (Brown et al. 2009, Andrews et al. 2011). The disk around HD 135344B is also known to show strong CO emission (Coulson et al. 1998, Thi et al. 2001, Dent et al. 2005), which provides an opportunity to investigate its molecular gas content and kinematics.

In this paper, we present comprehensive 1.3 millimeter observations of HD 135344B, in continuum and the $^{12}$CO, $^{13}$CO, and C$^{18}$O J=2–1 lines obtained with the Submillimeter Array (SMA)[1] using observations from multiple antenna configurations. These data allow us to confirm the inner cavity resolved by Brown et al. (2009) at 0.87 millimeters, and to provide the first information about the molecular gas disk size, geometry, and kinematics.

---

[1] The Submillimeter Array is a joint project between the Smithsonian Astrophysical Observatory and the Academia Sinica Institute of Astronomy and Astrophysics and is funded by the Smithsonian Institution and the Academia Sinica.

## 2. Observations and Data Reduction

The SMA (Ho, Moran & Lo 2004) was used to obtain four tracks of 1.3 millimeter data of HD 135344B in 2007 and 2008 using three configurations of the array, resulting in projected baselines from 7 to 510 meters. Table 1 lists the observational parameters. The total bandwidth available was 2 GHz in each of two sidebands. The spectral resolution was 0.20 MHz, for the $^{12}$CO and $^{13}$CO J=2–1 lines at 230.53797 GHz and 220.39868 GHz, respectively, corresponding to velocity resolution of ~0.27 km sec$^{-1}$. The C$^{18}$O J=2–1 line at 219.56036 GHz was also included in the setup, though with only a coarse spectral resolution of 3.2 MHz (~4.44 km sec$^{-1}$). The phase tracking center was specified as R.A. = $15^h15^m48^s.40$, DEC =$-37°09'16''.0$ (J2000). The field of view is ~51″ at these frequencies. Bandpass calibration was done using available bright quasars, J1751+096, 3C273 and 3C454.3, and flux calibration was done by observing the standard sources Uranus, Titan and Callisto. The flux calibration accuracy is approximately 10 percent. Complex gains were calibrated with interleaved observations of the nearby quasars J1517-243 and J1454-377. We calibrated the data using the MIR software package adopted for the SMA. Imaging was done using the miriad software.

For the continuum, the synthesized beam size using uniform weighting and all available visibilities is $1''.07 \times 0''.48$ with position angle of 19° and rms noise 1.0 mJy beam$^{-1}$. Using only the data obtained in the very extended configuration results in a smaller beam size of $0''.88 \times 0''.33$ at position angle 21°, with a higher rms noise of 1.7 mJy beam$^{-1}$. For the $^{12}$CO J=2–1 line, the synthesized beam size using natural weighting is $1''.16 \times 0''.65$ with position angle 19°, and rms noise 50 mJy beam$^{-1}$ in a 0.27 km sec$^{-1}$ velocity bin. For the $^{13}$CO J=2–1 line, the corresponding values are $1''.23 \times 0''.68$ at position angle 18° and rms noise 67 mJy beam$^{-1}$. For the C$^{18}$O J=2–1 line, we present an image using data just from the compact configuration, with synthesized beam size 4″.3 × 2″.5 at position angle 32° and rms noise 30 mJy beam$^{-1}$ in a 4.44 km sec$^{-1}$ velocity bin.

## 3. Results

### 3.1. Dust Continuum Emission

The 1.3 millimeter dust continuum emission was clearly detected toward HD 135344B. Figure 1 shows the continuum emission imaged in two ways: the left panel shows an image using all of the data with uniform weighting, and the right panel shows an image using data from just the very extended configuration, which provides a higher resolution and emphasizes smaller scale features.

The lower resolution image shows a structure elongated from northeast to southwest. Note that the elongation is significant even though the synthesized beam also has an elongation in a similar direction. From the lower resolution image, we find a total flux density of 141 mJy, consistent with the 142±19 mJy obtained at approximately the same wavelength by Sylvester et al. (1996) using the single dish James Clerk Maxwell Telescope. From this measurement, a crude estimate of the disk mass can be derived using the following formula (e.g. Thi et al. 2001 and references therein):

$$M_{\text{disk}} = 0.06 M_\odot \frac{F_\lambda}{1\text{Jy}} \left(\frac{d}{100\text{pc}}\right)^2 \frac{50\text{K}}{<T>} \frac{0.01 \text{cm}^2 \text{g}^{-1}}{\kappa_{1.3\text{mm}}} \qquad (1)$$

The $^{12}$CO J=2–1 emission suggests a gas temperature T ≤ 30 K (see §3.2), although the disk midplane where most of the dust is located may be colder than the CO emitting region. Taking this temperature and a dust opacity $\kappa_{1.3\text{mm}} = 0.01$ (including a gas to dust ratio of 100) (Ossenkopf & Henning 1994) gives a disk mass $M_{\text{disk}} = 2.8 \times 10^{-2} M_\odot$.

The higher resolution dust continuum image clearly shows a double-peaked morphology indicative of a central cavity resolved with an elliptical beam. This morphology is similar to that of 7 millimeter continuum emission from the disk around TW Hya observed with VLA (Hughes et al. 2007). The disk central position identified from the integrated $^{12}$CO J=2–1 emission (§3.2) is located at the middle of the two peaks. The separation of the two peaks (east and west) suggests an inner hole radius of 0″.36, or ~50 AU. This cavity size is consistent with that found previously by Brown et al. (2009). In addition, the asymmetric appearance of the disk in Figure 1, with brighter emission to the southeast, is consistent with the image at 0.87 millimeters of Brown et al. (2009).

## 3.2. CO Line Emission

The CO line emission provides direct information about the disk geometry and kinematics, and also constraints on physical properties of the molecular gas.

### *3.2.1. Geometry and kinematics of the gas disk*

Figure 2 shows channel maps of the $^{12}$CO J=2–1 line made with natural weighting. The $^{12}$CO J=2–1 line was detected at 3 sigma level over the LSR velocity range from 4.67 km s$^{-1}$ to 9.80 km s$^{-1}$. A clear velocity gradient from the north-east to the south-west can be discerned from the channel maps, suggestive of disk rotation. Figure 3 shows channel maps of the $^{13}$CO J=2–1 line, which was detected at 3 sigma level over a comparable LSR velocity range as the main isotope line. These channel maps also exhibit a velocity gradient from the north-east to the south-west, consistent with the $^{12}$CO J=2–1 line.

Figure 4 shows integrated intensity maps of $^{12}$CO (left panel) and $^{13}$CO (right panel) in contour, superposed on their intensity-weighted mean velocity maps in color. The $^{12}$CO integrated intensity map shows an elliptical structure elongated from the north-east to the south-west direction, in the same direction of the elongation of the continuum emission and the velocity gradient. The map shows a single peak near the center of the elliptical structure. This peak position is coincident with the dynamical center determined from the high velocity emission in the $^{12}$CO channel maps. We identify this peak position [R.A. = $15^h15^m48^s.44$, DEC $=-37°09'16".16$ (J2000)] as the disk center. Note that this disk center position is offset from the phase tracking center of the observations by ($0".47, -0".17$).

The $^{13}$CO integrated intensity map shown in Figure 4 also shows an elongated structure with a single peak, but the peak position is shifted from the disk center to west by ~$0".25$. The map also shows an extension from the center to south-east, making a double-peak morphology together with the western peak. This double-peak morphology is similar to the 1.3 millimeter continuum emission. Figure 5 shows a higher resolution $^{13}$CO J=2–1 image made with uniform weighting, superposed on the 1.3 millimeter continuum image. This higher resolution line image exhibits a double-peak morphology even more clearly, and suggests diminished emission in the inner gas disk. Careful comparison between the 1.3 millimeter continuum emission and the $^{12}$CO emission shows that the contrast between the south-east and north-west sides are different; the 1.3 mm continuum emission is stronger on the south-east side, while the $^{13}$CO emission shows the opposite tendency.

Figure 6 shows an image of the C$^{18}$O J=2–1 integrated emission, detected at the 4σ level. The signal is not strong enough to image at higher angular resolution. While this image does not show any evidence of a central hole like the $^{13}$CO J=2–1 line, in part because of the poor resolution, the emission peak does appear shifted slightly to the west, similar to the $^{13}$CO J=2–1 emission.

## 3.2.2. Physical properties of the gas disk

We can crudely estimate the optical depths of the $^{12}$CO and $^{13}$CO J=2–1 lines from an assumed isotopic line ratio,

$$R = \frac{T_R(^{12}CO)}{T_R(^{13}CO)} = \frac{1-e^{-\tau_{12CO}}}{1-e^{-\tau_{13CO}}} = \frac{1-e^{-\tau_{12CO}}}{1-e^{-\tau_{12CO}/X}} \tag{2}$$

where we take X =[$^{12}$CO]/[$^{13}$CO] = 60 (Wilson & Rood 1994). This gives $\tau_{12CO}$ ~30 and $\tau_{13CO}$ ~0.5 with a ratio R ~3 obtained from the comparison of the entire velocity range of the $^{12}$CO and $^{13}$CO J=2–1 emission. Since the $^{12}$CO line turns out to be optically thick, the observed intensity of this line provides an estimate of the gas temperature. The peak flux of 1.08 Jy beam$^{-1}$ suggests ~30 K, for the $^{12}$CO J=2–1 emission surface.

Assuming the $^{13}$CO J=2–1 line is optically thin, we can estimate the disk gas mass, from

$$M = N(total)_{13CO} \left[\frac{H_2}{^{12}CO}\right]\left[\frac{^{12}CO}{^{13}CO}\right]\mu m_{H_2}\Omega d^2 \tag{3}$$

where μ = 1.36, $m_{H_2}$ is the hydrogen molecular mass, d is the distance, Ω is the solid angle subtended by the $^{13}$CO molecular emission region, and N(total) is the total column density of $^{13}$CO, given by

$$N(total) = N(J)\frac{Z}{(2J+1)}\exp\left[\frac{hB_eJ(J+1)}{kT}\right] \tag{4}$$

with Partition function

$$Z = \sum_{J=0}^{\infty}(2J+1)\exp\left[\frac{hB_eJ(J+1)}{kT}\right] \tag{5}$$

where J is the angular momentum quantum number of the lower level. The column density in a level $l$ of $^{13}$CO molecule is calculated by

$$N_l = 1.93 \times 10^3 \frac{g_l \nu^2}{g_u A_{ul}} \int T_B dv \qquad (6)$$

where $g_u$ is the statistical weight of the upper level, $g_l$ is the statistical weight of the lower level, $A_{ul}$ is the Einstein A coefficient, $\nu$ is the frequency of the transition, and $\int T_B dv$ is the integrated intensity. Using $A_{21} = 5.0 \times 10^{-5}$ and the $^{13}$CO rotational constant Be of ~55 GHz, the integrated line intensity, $\int T_B dv =$ 11.67 K km s$^{-1}$ gives N(total)13CO $\approx 6.7 \times 10^{15}$ cm$^{-2}$. Assuming the [H$_2$]/[$^{12}$CO] = $10^4$ and the isotopic ratio above results in a disk gas mass of ~$3.8 \times 10^{-4}$ M$_\odot$.

The disk mass estimated from $^{13}$CO emission is only ~1 percent of that derived from the 1.3 mm continuum (~$2.8 \times 10^{-2}$ M$_\odot$). The larger disk mass derived from the dust continuum emission is more favorable to the relatively large mass accretion rate ($10^{-8.27}$ M$_\odot$ yr$^{-1}$; Garcia Lopez et al. 2006) in the disk. The discrepancy in the derived disk mass suggests global CO gas depletion in the disk. Such CO gas depletion in circumstellar disks was previously reported as well (e.g., Thi et al. 2001). However, the considerable uncertainties in this analysis as the maximum uncertainties in the dust opacity (a factor of 5) (Ossenkopf & Henning 1994), dust temperature (a factor of 2), gas-to-dust ratio (a factor of 2), and the [$^{12}$CO]/[$^{13}$CO] isotope ratio (a factor of 2) may also reduce this overall depletion.

## 4. Discussion

### 4.1. Dust disk structure

The dust disk structure previously has been modeled in detail using axisymmetric radiative transfer models (Brown et al. 2009, Andrews et al. 2011). We investigate the dust disk structure using a very simple toy model–face-on, axisymmetric, constant surface density, constant temperature, inner hole augmented by point sources to account for observed asymmetries.

We derive 1.3 millimeter continuum fluxes for the various components of the model with the miriad task uvfit. We fixed the outer radius at 90 AU derived from the uvfit to the combined all visibilities and an inner hole radius of 50 AU (see §3.1). The center position was fixed at 0″.47 east and 0″.17 south of the phase tracking center, based on the peak of the integrated $^{12}$CO J=2–1 emission. The color images on left-hand side of the 230 GHz data in Figure 7 show the image made from the toy-model described above, and the residual after the image made from the toy-model is subtracted from the observed image For comparison, the observed image is shown in contour. The total flux densities derived for the outer disk and inner hole regions are ~111 mJy and ~3 mJy, respectively. This suggests an averaged surface density contrast of ~20 between the outer disk and inner hole; any temperature gradient from stellar heating would increase this ratio. Figure 7 shows the data, this axisymmetric model, and the residuals obtained by imaging the data after subtracting the model. The most outstanding feature is a significant residual in the south-east part of the disk. This residual can be accounted for with two additional point-like components, i.e., a 14 mJy one at (0.48, -0.73) on the outer disk and a 6 mJy one at (0.68, -0.22) within the hole. As shown in Figure 7, this ad hoc composite model reduces the residuals significantly and reproduces the observed asymmetries well.

Consistency with independent observations builds confidence in the observed asymmetries. We apply the same toy model to the 0.87 millimeter dust continuum data obtained by Brown et al. (2009). The derived total flux densities of the outer disk and the inner hole regions are ~374 mJy and ~11 mJy, respectively, approximately 3 times higher than those estimated from the 1.3 millimeter data. In the right panels of the Figure 7 shows the data, model, and residuals in the same format as 230 GHz data. The residuals show a similar asymmetric feature in the south-east, and these may be reduced significantly by adding the same two point sources to the model, scaling their flux by a factor of 3 (*Flux* ~ $v^{\alpha}$ at $\alpha = 2+\beta$). The estimated value of $\beta=1$ suggested that the dust grain size is larger than in molecular clouds typically having $\beta \sim 2$ (Hildebrand 1983, Beckwith et al. 1990).

### 4.2. Gas disk structure

To obtain more information about the disk geometry and kinematics from the molecular line emission, we apply radiative transfer modeling and follow a procedure similar to that of Rodriguez et al. (2010). We assume a Keplerian disk in hydrostatic equilibrium with radial density and temperature described by truncated power laws, i.e. $n(r) = n_{100}(r/100\text{AU})^{-p}$ and $T(r) = T_{100}(r/100\text{AU})^{-q}$, where $n(r)$ refers to the number density of $H_2$, and we assume $[H_2]/[^{12}CO]=10^4$. There are 9 parameters that describe the Keplerian disk model (see Table 2), and the modeling can constrain four: the CO disk outer truncation radius ($R_{out}$), the CO disk inclination (i), the systemic velocity ($V_{LSR}$) and the disk position angle. The modeling is relatively insensitive to the other parameters because the $^{12}CO$ J=2-1 emission is optically thick. We fixed the values of the stellar mass, $T_{100}$, q, $n_{100}$ and p and varied the remaining four model parameters in turn ($R_{out}$, i, $V_{LSR}$, P.A.) to determine the best-fit values. The uncertainties were determined from the change in $\chi^2$ as the model parameters were varied and they did not account for systematic errors. Similar analysis has been carried out in Rodriguez et al. (2010). Table 2 lists these best-fit values and their formal 1σ uncertainties. The fixed values related to the temperature and density distributions are q=0.8, $T_{100}=26$ K and p=2.5, $n_{100}=1.7 \times 10^9$ cm$^{-3}$. The derived best-fit values of the disk size, inclination, systemic velocity and position angle are $R_{out}=220\pm20$ AU, i=11.5±0.5°, $V_{LSR}=7.12\pm0.02$ km s$^{-1}$ and P.A.=64±2°, respectively. The derived CO disk outer radius is much larger than the dust outer disk size of $R_{out}=90$ AU.

Similar difference in the disk size between CO and dust is also seen in other circumstellar disks, such as HD163296 (Isella et al. 2007) and AB Aur (Piétu et al. 2005, Lin et al. 2006). Such difference in the disk size is often interpreted as difference in the optical depth, but not as real difference in the disk size, i.e, a lower optical depth of dust continuum emission as compared to that of molecular line emission makes the appearance of dust disks smaller than gas disks (Dutrey et al. 2007, Andrews et al. 2009).

Figure 8 compares the observed channel maps of the $^{12}$CO J=2–1 line and channel maps made from the best-fit model, and shows the residuals between the observations and the model (binned to 0.6 km s$^{-1}$ resolution for clarity). The derived systemic velocity of 7.1 km s$^{-1}$ is consistent with the result from the $^{12}$CO J=3–2 single dish observation by Thi et al. (2001). The disk inclination is similar to the value of 11±2° obtained by Dent et al. (2005) from modeling the $^{12}$CO J=3–2 single dish line profile. The disk size, however, is much larger than the Rout ~125 AU (D$_{out}$=1.8 arcsec) obtained by Dent et al. (2005) without any spatially resolved data.

The $^{13}$CO J=2–1 integrated emission map shows a double-peak morphology that suggests that $^{13}$CO (and possibly $^{12}$CO) gas may be depleted in the inner disk, like the dust. To investigate the contrast between the inner and outer gas disk, we modeled the $^{13}$CO J=2–1 emission using several different assumptions. Figure 9 presents simulated channel maps for four different models of the $^{13}$CO J=2–1 emission. Note here we use only the extended and very extended configuration data to achieve a higher resolution and retain good brightness sensitivity. Model 1 assumes that the $^{13}$CO J=2–1 distribution follows the best-fit model of $^{12}$CO J=2–1, but the density and outer radius are varied to fit the $^{13}$CO data. Model 1A and 1B are similar to Model 1 except the density of $^{13}$CO is made lower within 50 AU radius by factors of 10 and 20, respectively. Model 1C has a completely empty hole of 50 AU radius. The lower panel of Figure 9 shows the integrated intensity maps of the $^{13}$CO J=2–1 data and for these four models. All of the models except Model 1 show the clear double-peak morphology. However, the $^{13}$CO J=2–1 channel maps clearly show emission at the higher redshifted velocities (8.92 to 9.52 km s$^{-1}$), indicating $^{13}$CO is not completely depleted in the inner 50 AU. These models cannot uniquely constrain the details of the gas structure, and they make no attempt to account for asymmetries, but they do strongly suggest an order of magnitude reduction in gas density in the inner disk, where the dust emission shows a clear cavity.

### 4.3. Origin of the Central Cavity

The main models for creating central cavities in disks involve dynamical clearing by companions or planets, evaporation by high energy photons, and grain growth that dramatically lowers the dust opacity. In the case of HD 135344B, however, dynamical clearing by companions or planets seems to be ruled out because of the null results of the several previous companion-searching studies in HD 135344B (Pontoppidan et al. 2008, Grady et al. 2009, Müller et al. 2011). In addition, the fact that the gas emission as well as the dust emission is reduced in the central cavity suggests that grain growth alone is not a viable mechanism to create the gap in this system.

The photoevaporation by energetic radiation (far-UV (FUV); 6 eV <hν < 13.6 eV, extreme-UV (EUV); 13.6 eV <hν < 0.1 keV and X-rays; hν > 0.1 keV) can be considered because it has been proposed as an important mechanism for the disk dissipation and gap formation (Hollenbach et al. 1994, Clarke et al. 2001). Recently, Gorti & Hollenbach (2009) showed that photoevaporation with UV and X-ray can create a large gap from ~7 AU to ~55 AU in radius around a 1.7 M$_\odot$ central star when the mass accretion rate is $10^{-8.27}$ M$_\odot$ yr$^{-1}$ (see their Fig 7 in details). In this case, however, the disk lifetime is estimated to be ~2 Myr, which is much shorter than the age of 8-17 Myr for HD 135344B. They might have adopted much larger UV and X-ray radiations in their model than the real environments (see also Andrews et al. 2011).

Considering the large disk mass of ~0.028 $M_\odot$ derived from the 1.3 millimeter continuum data and the high mass accretion rate of $\dot{M}=10^{-8.27} M_\odot$ yr$^{-1}$ of this disk system, we suggest that the inner cavity detected in the disk is more plausibly related to planet formation than to photoevaporation (Andrews et al. 2011; see also, Williams and Cieza 2011). Furthermore, the ~10 times lower mass accretion rate compared to that of the non-transition CTT stars at the same disk mass supports the prediction for the Jovian-mass planet formation scenario (Lubow & D'Angelo 2006, Najita et al. 2007).

The observed asymmetric appearance of the dust continuum emission also may be related to planet formation in this disk system. The azimuthal variations (asymmetric features as discussed in §4.1) in the disk systems may be explained by either the disk self-gravity or proto-planets (Williams & Cieza 2011). In this case, the disk self-gravity must be negligible because the kinematics are consistent with Keplerian rotation, indicated by the resolved spatial and velocity information of the $^{12}$CO and $^{13}$CO J=2-1 lines (§3.2.1 and §4.2).

## 5. Summary

We present high resolution (~1″) observations with the SMA of the ~8–17 Myr-old transitional disk around the F star HD 135344B in the 1.3 millimeter dust continuum and the CO J=2–1 line and isotopes. The high-spatial resolution results provide direct information about the structure, physical properties, and kinematics of the disk, which provide clues to further test theoretical mechanisms of disk clearing. The main results are summarized as follows:

1. We find an inner cavity in the 1.3 millimeter dust emission with radius R ~50 AU, in accord with previous observations at 0.87 millimeters. The emission shows a significant asymmetry, with a brighter feature in the south-east part of the disk, again similar to previous observations. A simple toy model of a disk with ~20 times lower column density in the central cavity and two point sources in the south-east can reproduce the observations.

2. The resolved spatial and velocity information from the $^{12}$CO and $^{13}$CO 2-1 lines show that the disk follows Keplerian rotation. The disk inclination is i ~11° and the position angle is ~64°. The $^{13}$CO 2-1 emission shows a central depression of size comparable to the dust continuum cavity. This feature can be modeled as the result of an order of magnitude reduction in gas column density within a radius of R ~50 AU.

3. We obtained a disk mass from the 1.3 mm continuum emission of ~$2.8 \times 10^{-2}$ $M_\odot$, about two orders of magnitude larger than that from the $^{13}$CO emission of ~$3.8 \times 10^{-4}$ $M_\odot$, which suggests an overall CO gas depletion in the disk, although the uncertainties in these mass estimates are considerable.

4. We suggest that the central cavity in gas and dust is more plausibly related to planet formation than to photoevaporation or to grain growth, given the high disk mass and mass accretion rate, and the reduction of both dust and gas in the inner disk.

Table 1: HD 135344B Observational Parameters.

| Parameter | compact-north | extended | extended | very-extended |
|---|---|---|---|---|
| Date of observation | 2007 Apr 9 | 2008 Jan 24 | 2008 Feb 12 | 2008 Apr 6 |
| Number of antennas | 8 | 5 | 7 | 7 |
| Projected baseline range (m) | 7–113 | 42–180 | 12–180 | 35–512 |
| Flux calibrator | Uranus & Callisto | Titan | Titan | Callisto |
| Gain calibrator | J1517-243 & J1454-377 | J1454-377 | J1454-377 | J1454-377 |
| Passband calibrator | J1751+096 | 3C273 | 3C273 | 3C454.3 |

Table 2: Best-Fit Model Parameters

| Parameter | CO 2-1 |
|---|---|
| Central Mass ($M_\odot$) | 1.8 |
| $T_{100}$ (K) | 26 |
| q | 0.8 |
| $n_{100}$ (cm$^{-3}$) | $1.7 \times 10^9$ |
| p | 2.5 |
| $R_{out}$ (AU) | 220±20 |
| i (°) | 11.5±0.5 |
| $V_{LSR}$ (km s$^{-1}$) | 7.12±0.02 |
| P.A. (°) | 64±2 |

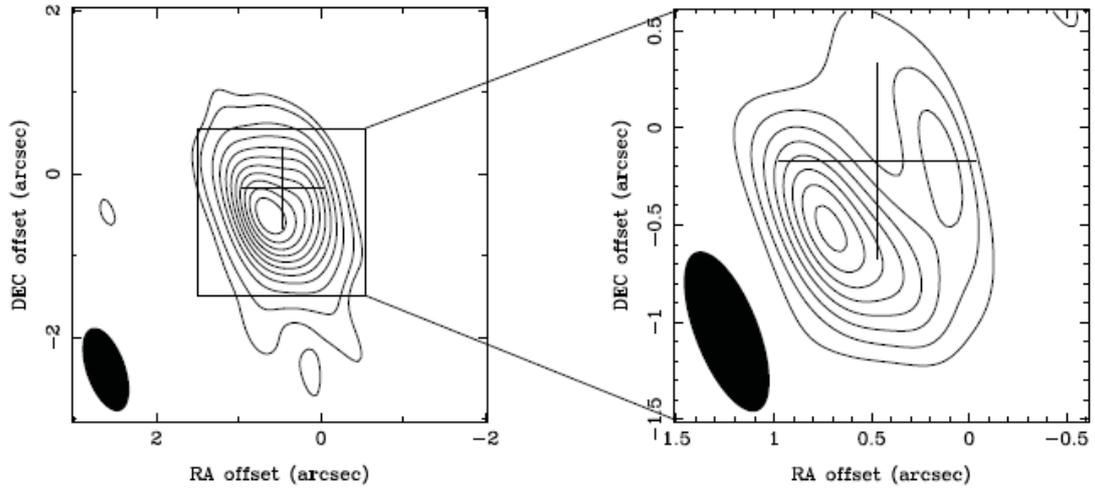

Fig. 1.— SMA 1.3 millimeter continuum maps of HD 135344B. The left panel shows an image obtained from the four tracks. The contour levels are 3, 5, 10, 15, 20, 25, 30, 35, 40, 45 and 50 ×1.0 mJy beam$^{-1}$, the rms noise level. The ellipse in the lower left corner represents the FWHM size of the synthesized beam, 1″.07 × 0″.48, P.A. = 19°. The right panel shows a higher resolution image obtained only from the array configuration with the longest baselines. The contour levels are 3, 5, 7, 9, 11, 13, 15 and 17 ×1.7 mJy beam$^{-1}$, and the FWHM size of the synthesized beam is 0″.88 × 0″.33, P.A. = 21°. The cross in each panel indicates the center position of the integrated $^{12}$CO J=2–1 emission.

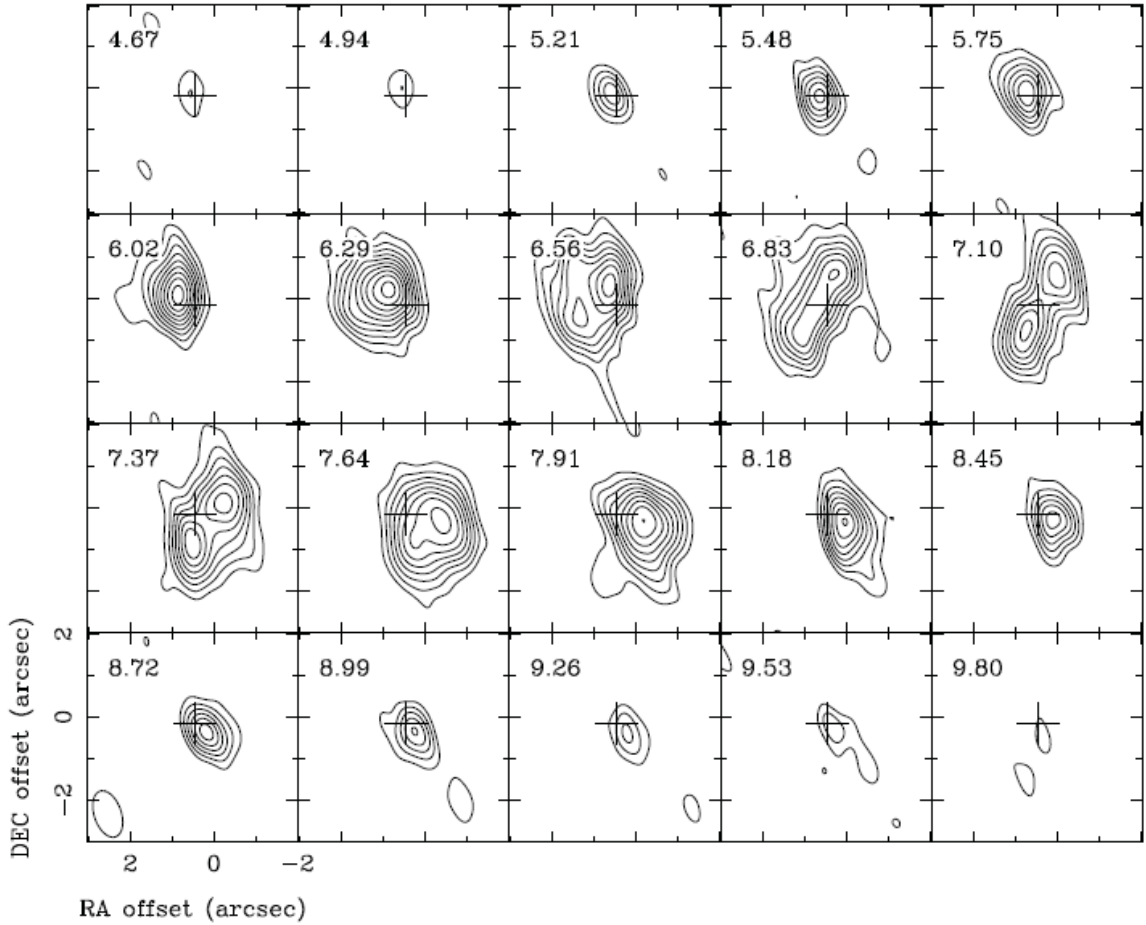

Fig. 2.— $^{12}$CO J=2–1 channel maps from the observations from all four configurations with natural weighting. The velocity of each channel is shown in the upper left-hand corner of each panels. The systemic velocity of HD 135344B is ~7.1 km sec$^{-1}$ and the channel velocity resolution is 0.27 km sec$^{-1}$. The contour levels are 3, 5, 7, 9, 11, 13, 15, 17 and 19 × σ, where σ = 0.05 Jy beam$^{-1}$. The synthesized beam is shown in the lower left-hand corner of the first panel, with FWHM size 1″.16 × 0″.65, P.A. 19°. The cross represents the center position of the integrated $^{12}$CO J=2–1 emission map.

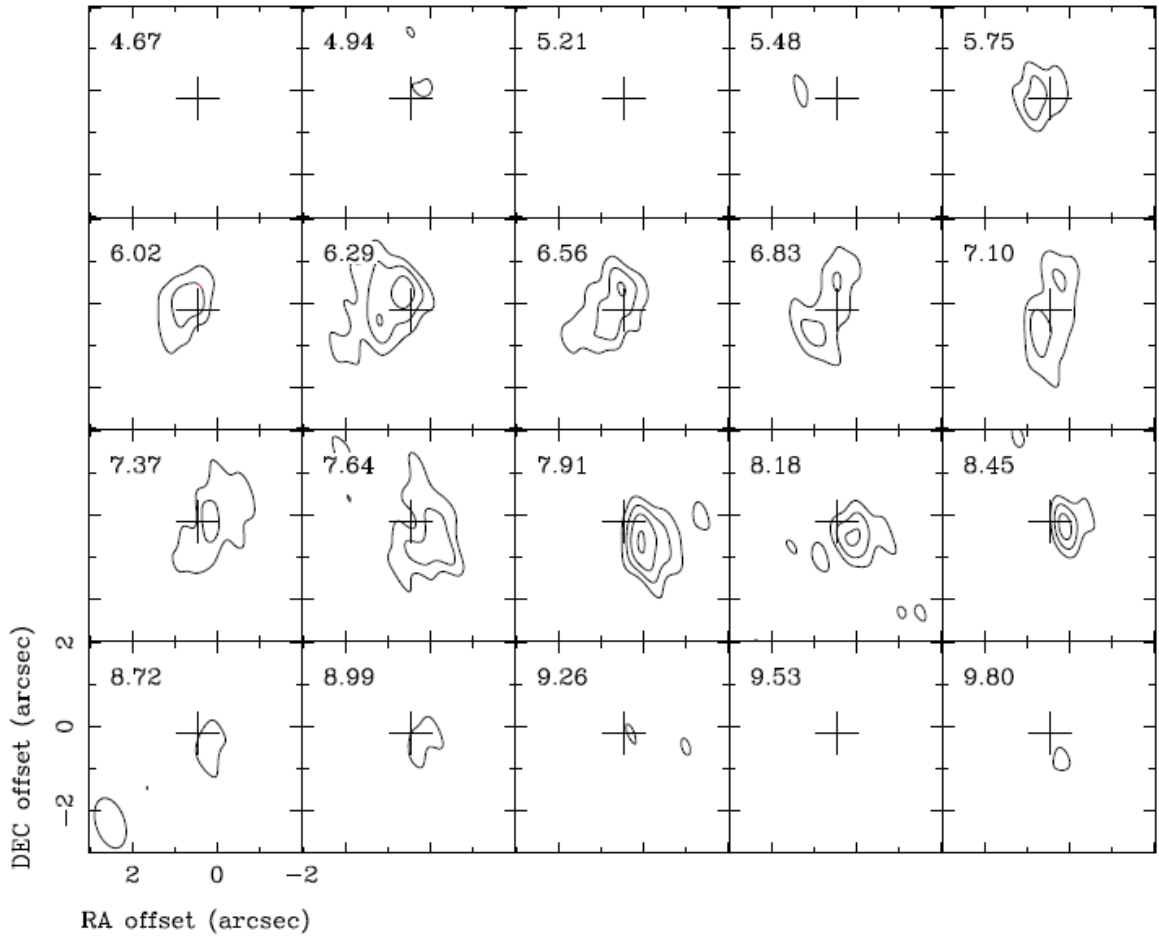

Fig. 3.— $^{13}$CO J=2–1 channel maps from the observations from all four configurations with natural weighting. The velocity of each channel is shown in the upper left-hand corner of each panels. The contour levels are 3, 5, 7, 9 × σ, where σ = 0.067 Jy beam$^{-1}$. The synthesized beam is shown in the lower left-hand corner of the first panel, with FWHM size 1″.23 × 0″.68, P.A. 18°.

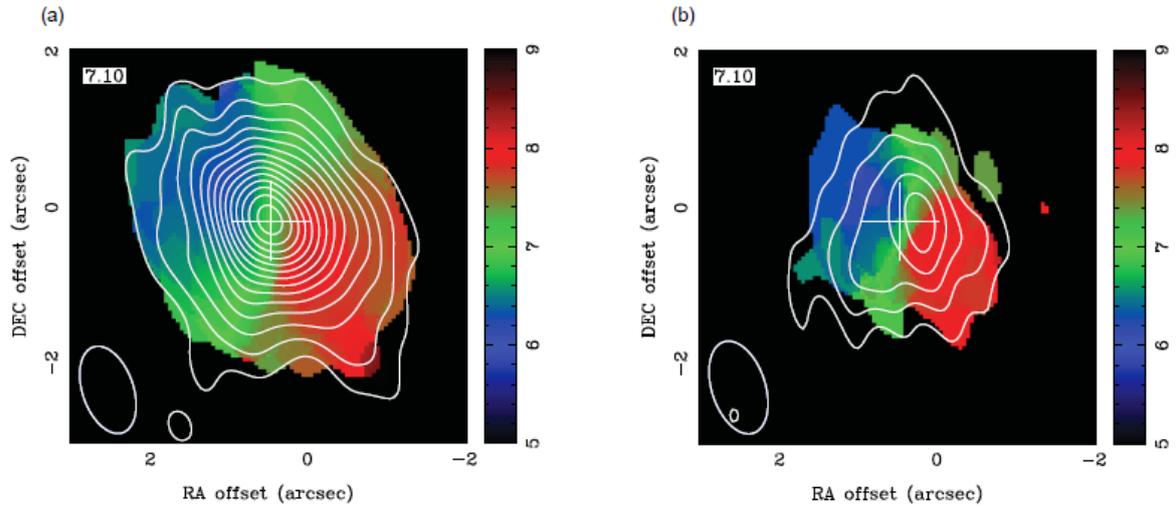

Fig 4.— Moment maps of $^{12}$CO J=2–1 and $^{13}$CO J=2–1. The cross in each panel represents the center position of the integrated $^{12}$CO J=2–1 emission map. (a) Contour plots are for the integrated $^{12}$CO J=2–1 emission over the entire velocity range, 5–9 km sec$^{-1}$. The contour levels are 3, 5, 7, 9, 11, 13, 15, 17, 19, 21, 23, 25, 27 and 29 × σ, where σ = 0.08 Jy beam$^{-1}$ km sec$^{-1}$. The color scales represent the intensity weighted velocity map. The blue-and red-shifted emission from the systemic velocity, ~7.1 km sec$^{-1}$, are shown in their colors. The synthesized beam is shown in the lower-left corner. (b) Contour plots are for the integrated $^{13}$CO J=2–1 emission over the entire velocity range, 5–9 km sec$^{-1}$. The contour levels are 3, 5, 7, 9 and 11 × σ, where σ = 0.11 Jy beam$^{-1}$ km sec$^{-1}$. The color scales represent the intensity weighted velocity map. The blue- and red-shifted emission from the systemic velocity, ~7.1 km s$^{-1}$, are shown in their colors. The synthesized beam is shown in the lower-left corner.

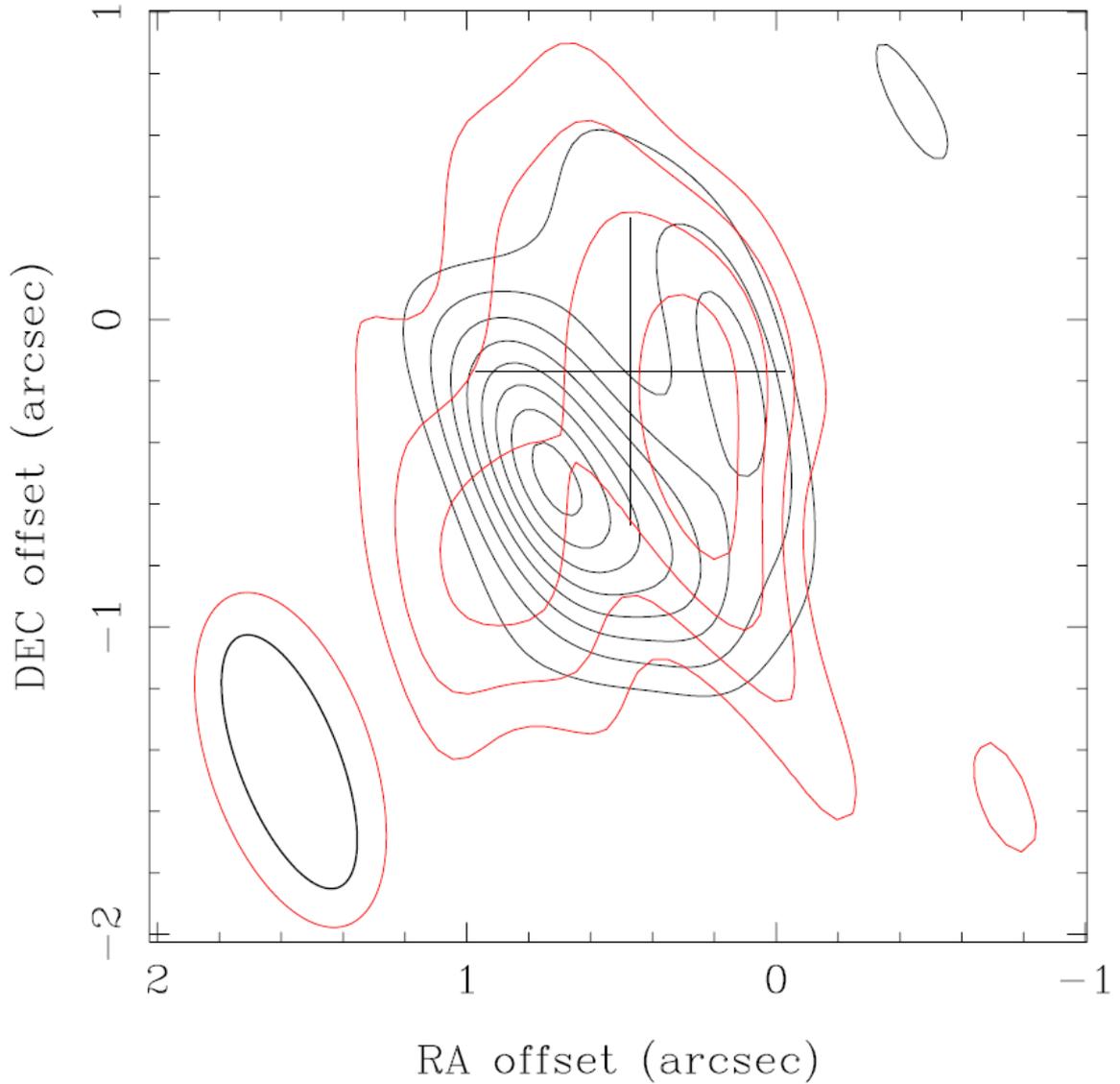

Fig. 5.— The black-solid lines represent the high-resolution 1.3 millimeter continuum map, with beam size FWHM of $0''.88 \times 0''.33$, P.A. 21°). The red-solid lines are for the integrated $^{13}$CO J=2–1 emission over the entire velocity range with the uniform weighting, with beam size FWHM of $1''.14 \times 0''.53$, P.A. 19°). The contour levels for dust continuum are 3, 5, 7, 9, 11, 13, 15 and 17 × σ at σ = 1.7 mJy beam$^{-1}$. The contour levels for the integrated $^{13}$CO emission are 3, 4, 5 and 6 × σ, where σ = 0.174 Jy beam$^{-1}$ km sec$^{-1}$.

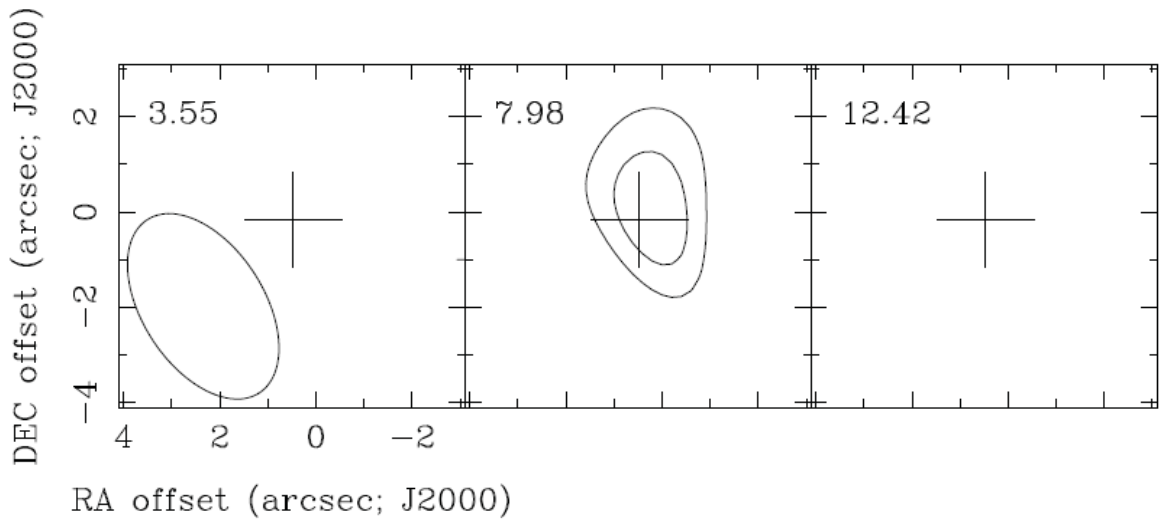

Fig. 6.— C$^{18}$O J=2–1 channel maps of the compact-north configuration data with natural weighting. The velocity of each channel is shown in the upper left-hand corner of each panel with the channel velocity resolution, 4.44 km sec$^{-1}$. The contour levels are 3 and 4 × σ, where σ = 0.03 Jy beam$^{-1}$. The synthesized beam is shown in the lower left-hand corner of the first panel, FWHM 4″.3 × 2″.5, P.A. of 32°. The cross is the center position of the integrated $^{12}$CO J=2–1 emission map.

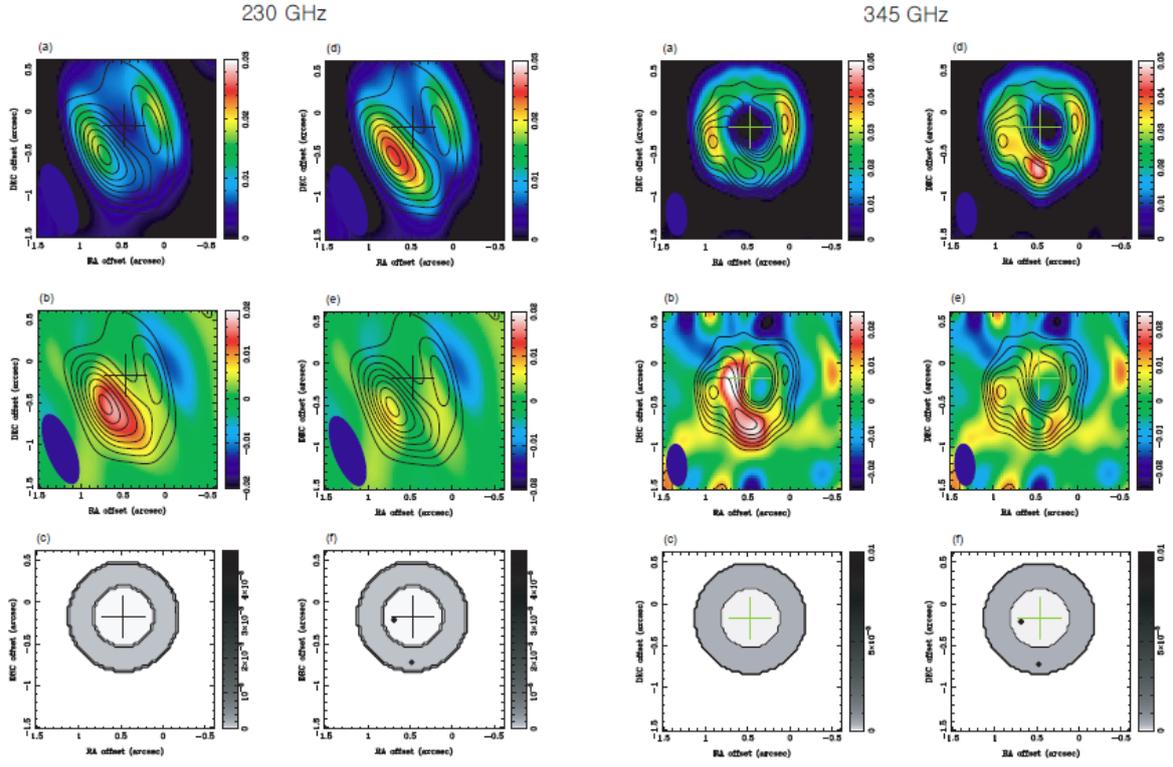

Fig. 7.— Toy-disk models for 1.3 mm continuum emission (in the left six panels) and 0.87 mm continuum emission (in the right six panels; the observational data from Brown et al. 2009). In the both sides, contours represent the observed images. In the left-hand side for 1.3 mm continuum map, the FWHM of the synthesized beam and the contour levels are same as those in Fig. 1. In the right-hand side for 0.87 mm continuum map, the FWHM of the synthesized beam is 0″.50 × 0″.24, P.A. of 5° and the contours start at 2σ with 1σ contour steps (1σ = 6.0 mJy beam$^{-1}$). In the both sides, (a) The color scale shows the simulated image of the toy-disk model(c). (b) The color scale represents the difference between the observational data and the toy-disk model(c). (c) The toy-disk model comprised of a face-on, axisymmetric, constant surface density disk of 90 AU radius with an inner hole of 50 AU radius having ~20 times lower surface density compared with that of the outer disk. (d) The color scale shows the simulated image of the toy-disk model(f). (e) The color scales represents the difference between the data and the toy-disk model(f). (f) The toy-disk model(c) augmented with two additional point sources.

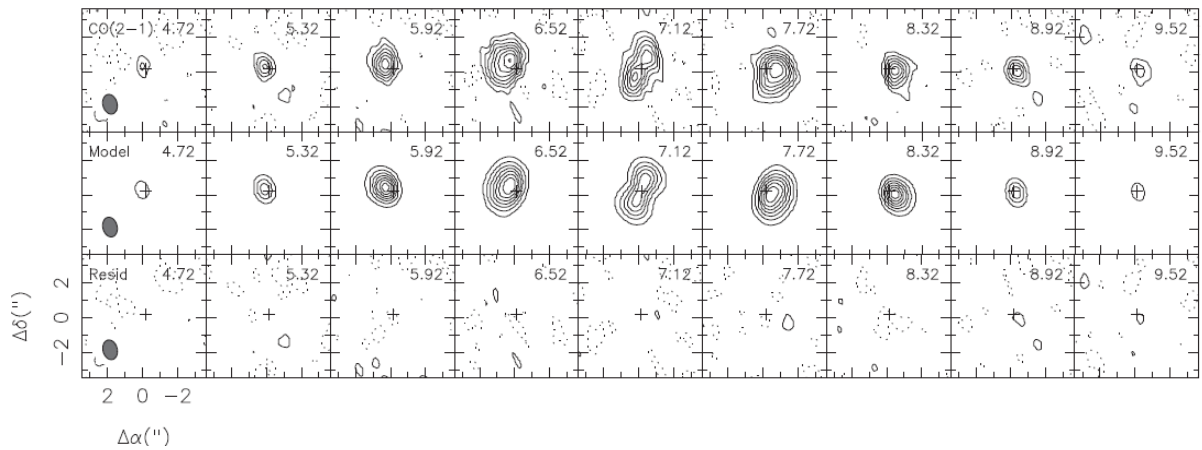

Fig. 8.— Top: Velocity channel maps of the $^{12}$CO J=2–1 emission toward HD 135344B. The angular resolution is 1″.1 × 0″.8, P.A. 13.9°. The axes are offsets from the pointing center in arcseconds. The contours start at 2σ with 2σ contour steps (1σ=0.07 Jy beam$^{-1}$). Middle: channel map of the best-fit model with the same contour levels. Bottom: difference between the best-fit model and data on the same contour scale.

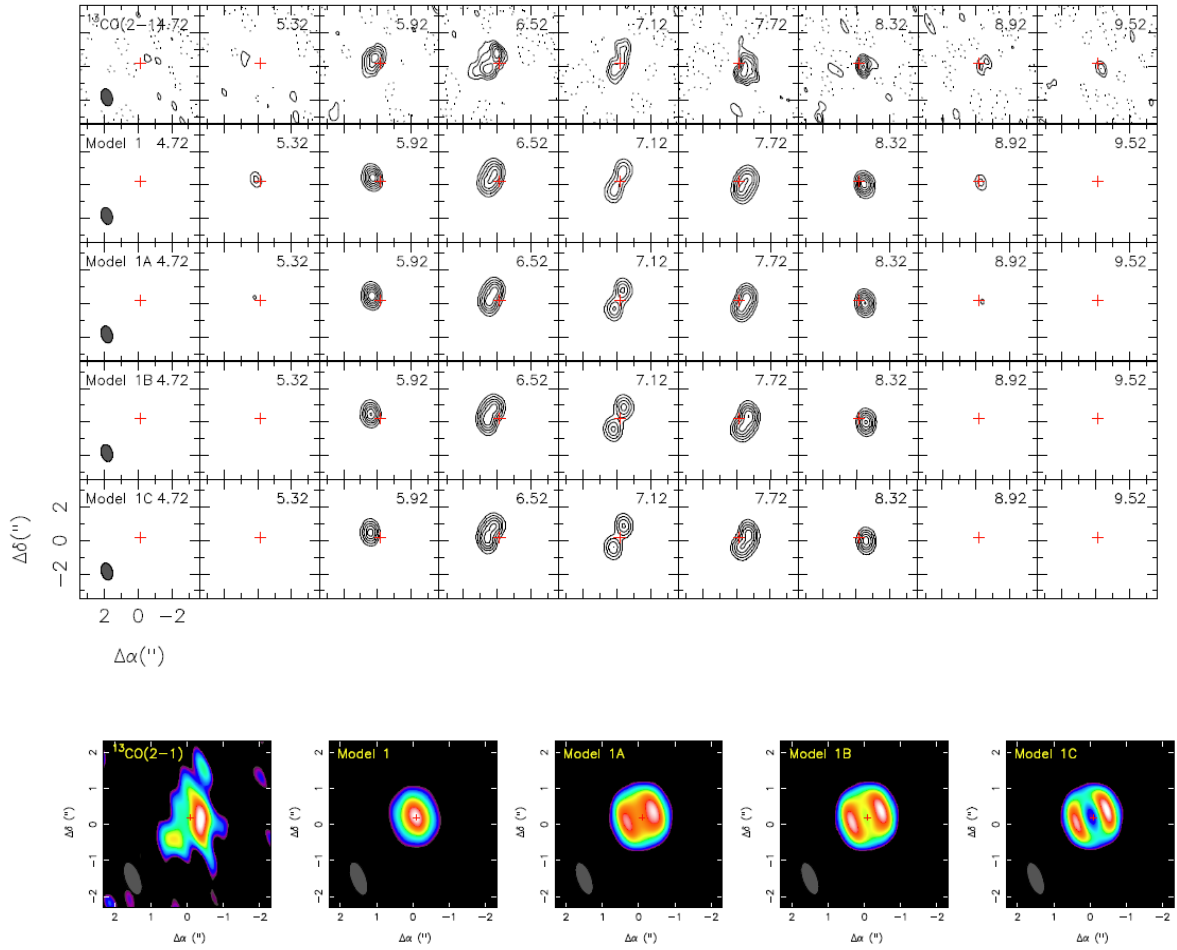

Fig. 9.— Upper panel: Velocity channel maps of the $^{13}$CO J=2–1 emission toward HD 135344B and the simulated models. The angular resolution is $1''.0 \times 0''.7$ at PA 14.6°. The axes are offsets from the pointing center in arcseconds. The $1\sigma$ contour step is 0.06 Jy beam$^{-1}$ and the contours start at $2\sigma$. Lower plane: the integrated intensity maps of the data and models.

# REFERENCES

Andrews, S. M., Wilner, D. J., Hughes, A. M., Qi, Chunhua, Dullemond, C. P. 2009, ApJ, 700, 1502

Andrews, S. M., Wilner, D. J., Espaillat, C., Hughes, A. M., Dullemond, C. P., McClure, M. K., Qi, C., Brown, J. M. 2011, ApJ, 732, 42

Beckwith, S. V. W., Sargent, A. I., Chini, R. S., Guesten, R. 1990, AJ, 99, 924

Brown, J. M., Blake, G. A., Dullemond, C. P., Merin, B., Augereau, J. C., Boogert, A. C. A., Evans, N. J., II, Geers, V. C., Lahuis, F., Kessler-Silacci, J. E., Pontoppidan, K. M., van Dishoeck, E. F. 2007, ApJ, 664, L107

Brown, J. M., Blake, G. A., Qi, C., Dullemond, C. P., Wilner, D. J., Williams, J. P. 2009, ApJ, 704, 496

Clarke, C. J., Gendrin, A., Sotomayor, M. 2001, MNRAS, 328, 485

Coulson, I. M., Walther, D. M., Dent, W. R. F. 1998, MNRAS, 296, 934

Dent, W. R. F., Greaves, J. S., Coulson, I. M. 2005, MNRAS, 359, 663

Doucet, C., Pantin, E., Lagage, P. O., Dullemond, C. P. 2006, A&A, 460, 117

Dutrey, A., Henning, T., Guilloteau, S., Semenov, D., Pietu, V., Schreyer, K., Bacmann, A., Launhardt, R., Pety, J., Gueth, F. 2007, A&A, 464, 615

Fedele, D., van den Ancker, M. E., Acke, B., van der Plas, G., van Boekel, R., Wittkowski, M., Henning, Th., Bouwman, J., Meeus, G., Rafanelli, P. 2008, A&A, 491, 809

Garcia Lopez, R., Natta, A., Testi, L., Habart, E. 2006, A&A, 459, 837

Gorti, U., Hollenbach, D. 2009, ApJ, 690, 1539

Grady, C. A., Schneider, G., Sitko, M. L., Williger, G. M., Hamaguchi, K., Brittain, S. D., Ablordeppey, K., Apai, D., Beerman, L., Carpenter, W. J., Collins, K. A., Fukagawa, M., Hammel, H. B., Henning, Th., Hines, D., Kimes, R., Lynch, D. K., Menard, F., Pearson, R., Russell, R. W., Silverstone, M., Smith, P. S., Troutman, M., Wilner, D.,

Woodgate, B., Clampin, M. 2009, ApJ, 699, 1822

Hildebrand, R. H. 1983, R. Astron. Soc. 24, 267

Ho, Paul T. P., Moran, James M., Lo, Kwok Yung. 2004, ApJ, 616, L1

Hollenbach, D., Johnstone, D., Lizano, S., Shu, F. 1994, ApJ, 428, 654

Hughes, A. M., Wilner, D. J., Calvet, N., D'Alessio, P., Claussen, M. J., Hogerheijde, M. R. 2007, ApJ, 664, 536

Isella, A., Testi, L., Natta, A., Neri, R., Wilner, D., Qi, C. 2007, A&A, 469, 213

Lin, S-Y., Ohashi, N., Lim, J., Ho, P. T. P., Fukagawa, M., Tamura, M. 2006, ApJ, 645, 1297

Lubow, S. H., D'Angelo, G. 2006, ApJ, 641, 526

Mason, Brian D., Wycoff, Gary L., Hartkopf, William I., Douglass, Geoffrey G., Worley, Charles E. 2001, AJ, 122, 3466

Müller, A., van den Ancker, M. E., Launhardt, R., Pott, J. U., Fedele, D., Henning, Th. 2011, A&A, 530, 85

Najita, J. R., Strom, S. E., Muzerolle, J. 2007, MNRAS, 378, 369

Ossenkopf, V., Henning, Th. 1994, A&A, 279, 577

Papaloizou, J. C. B., Nelson, R. P., Kley, W., Masset, F. S., Artymowicz, P. 2007, in
Reipurth, B., Jewitt, D., Keil, K., eds, Protostars and Planets V, University of Arizona, p655

Piétu, V., Guilloteau, S., Dutrey, A. 2005, A&A, 443, 945

Pontoppidan, K. M., Blake, G. A., van Dishoeck, E. F., Smette, A., Ireland, M. J., Brown, J. 2008, ApJ, 684, 1323

Rodriguez, D. R., Kastner, J. H., Wilner, D. J., Qi, C. 2010, ApJ, 720, 1684

Strom, K. M., Strom, S. E., Edwards, S., Cabrit, S., Skrutskie, M. F. 1989, AJ, 97, 1451

Sylvester, R. J., Skinner, C. J., Barlow, M. J., Mannings, V. 1996, MNRAS, 279, 915

Thi, W. F., van Dishoeck, E. F., Blake, G. A., van Zadelhoff, G. J., Horn, J., Becklin,

E. E., Mannings, V., Sargent, A. I., van den Ancker, M. E., Natta, A., Kessler,

J. 2001, ApJ, 561, 1074

van Boekel, R., Min, M.; Waters, L. B. F. M., de Koter, A., Dominik, C., van den Ancker, M.

E., Bouwman, J. 2005, A&A, 437, 189

Williams, J. P., Cieza, L. A. 2011, astro-ph/1103.0556

Wilson, T. L., Rood, R. 1994, ARA&A, 32, 191